# RANGE ANXIETY AMONG BATTERY ELECTRIC VEHICLE USERS: BOTH DISTANCE AND WAITING TIME MATTER

Jiyao Wang[1], Chunxi Huang[2], Dengbo He[1,3], Ran Tu[4]

1. Systems Hub, The Hong Kong University of Science and Technology (Guangzhou), Guangzhou, China
2. Interdisciplinary Programs Office, The Hong Kong University of Science and Technology, Hong Kong SAR, China
3. Department of Civil and Environmental Engineering, The Hong Kong University of Science and Technology, Hong Kong SAR, China
4. School of Transportation, Southeast University, Nanjing, China

Range anxiety is a major concern of battery electric vehicles (BEVs) users or potential users. Previous work has explored the influential factors of distance-related range anxiety. However, time-related range anxiety has rarely been explored. The time cost when charging or waiting to charge the BEVs can negatively impact BEV users' experience. As a preliminary attempt, this survey study investigated time-related anxiety by observing BEV users' charging decisions in scenarios when both battery level and time cost are of concern. We collected and analyzed responses from 217 BEV users in mainland China. The results revealed that time-related anxiety exists and could affect users' charging decisions. Further, users' charging decisions can be a result of the trade-off between distance-related and time-related anxiety, and can be moderated by several external factors (e.g., regions and individual differences). The findings can support the optimization of charge station distribution and EV charge recommendation algorithms.

## INTRODUCTION

With the rapid development of battery technology, battery electric vehicles (BEVs) are considered a promising solution for vehicle fuel shortage and emission issues (Hannan et al., 2017). In the year 2019, the sales of electric vehicles reached a total of 2.1 million, exhibiting a yearly growth rate of 40% (Shibl et al., 2021). According to the sales report published by the Chinese government (Phyllis et al., 2022) in 2020, new energy vehicles are expected to occupy more than 50% of total vehicle sales in 2035, and BEVs will account for over 95% of it. However, to date, range anxiety is still one of the major obstacles to the popularization of BEVs with the current level of the battery and charging technologies (Franke et al., 2016). According to Rauh et al. (2015), range anxiety manifests as the driver's uncertainty about reaching the destination with the remaining battery capacity.

In order to alleviate range anxiety, previous research tried to optimize the charging station distribution (Bulut & Kisacikoglu, 2017; Hafez & Bhattacharya, 2017; L. Pan et al., 2020). However, upgrading or relocating infrastructure is costly. Thus, researchers also tried to investigate the socio-psychological factors leading to range anxiety. For example, researchers found that drivers' trust in the range estimation systems (RESs) (Hariharan et al., 2022; Wang et al., 2021) and their comfort range (i.e., the users' range comfort zone or range safety buffer) (Yuan et al., 2018) are dominating predictors of range anxiety. Rauh et al. (2015) found that the experience of using BEV is negatively associated with range anxiety level. The authors further explained that the experience with similar situations in BEVs allows drivers to come up with more potential solutions to the current situation, thus alleviating their anxiety about range uncertainty.

However, the research so far mostly focused on the range anxiety caused by "distance uncertainty"; while the "time uncertainty" might be of greater concern for BEV users at this stage. Time uncertainty can lead to time-related range anxiety. Following the definition of distance-related anxiety in Rauh et al. (2015), we define time-related range anxiety as the driver's uncertainty about reaching the destination in time. For example, most of the rest areas in the east part of China along the highway have been equipped with charging stations (The Beijing News, 2022). However, the time cost for recharging a BEV can be frustrating – in holidays, the queuing time can be as long as four hours (News China, 2022). Further, in China, the average range of BEVs sold in 2022 was 359 km (Phyllis et al., 2022); In contrast, in 2021, for 24 major cities in China, the average density of charging facilities exceeded 21.5 stations/km$^2$ (Chen Zhang, 2022), or 0.22 km between two stations in average, which is far less than the range of BEVs. However, range anxiety still troubles BEV users, potentially because of the increasingly fast growth of BEV user population and as a result, the long waiting time at the stations. Therefore, the benefits of increasing the density of charging stations in order to reduce distance-related anxiety might be limited at this stage, and time-cost-related factors may contribute more to BEV users' range anxiety. Actually, researchers have found that the time cost of charging BEVs can reduce potential users' intention to purchase BEVs (Zhang et al., 2021).

Although more advanced technologies (Chakraborty et al., 2022) have been utilized to speed up the charging time of BEVs, restricted by the infrastructure development and battery technology, BEVs in the market usually take from 0.5 to 1 hour to be fully recharged on average (Pod Point, 2021). Further, as the progress of technology is usually at a gradual and uncertain pace, and the costs for upgrading BEV hardware and charging infrastructure are high (Madina et al., 2016), the charging speed for BEVs may not improve significantly shortly. Thus, to better model drivers' energy replenishment decisions, and explore influential factors of heterogeneous range anxiety among BEV users, a new model that takes both distance anxiety and time anxiety into consideration is needed.

Hence, in this current study, as a preliminary work to model both distance-related and time-related range anxiety, an online questionnaire was designed and issued in mainland China to investigate factors influencing range anxiety. Being different from previous range anxiety studies, the questionnaire targeted scenarios where both "distance anxiety" and "time anxiety" may affect users' charging decisions. Factors affecting users' decisions in these scenarios were also assessed. The findings can guide customized charging recommendation and optimize charging station planning.

Table 1. Questionnaire Design and Variable Extraction

| Questions | Variables | Distribution of extracted variables |
|---|---|---|
| Q1: [FI] Date of birth. | Age | - Mean: 26.9 years old (SD: 5.4, min: 18, max: 48) |
| Q2: [SC] How frequently do your drive BEVs? | *Driving Frequency* | - Frequently (over once a week, n=187, 86.2%)<br>- Infrequently (n=30, 13.8%) |
| Q3: [FI] Please indicate the province you drive the most.<br>- Further categorized into three levels based on EV infrastructure development (Cheng et al., 2021). | *Infrastructure* | - Well developed (n=96, 44.2%)<br>- Average (n=86, 39.6%)<br>- Less developed (n=35, 16.2%) |
| Q4: [LS] Trustworthiness scale (FIFT) (Franke et al., 2015) regarding users' trust in RES of the BEV they drive the most.<br>- 1 ("not at all") to 7 ("extremely") | *BEV Trust* | - Mean: 5.87 (SD: 1.1, min: 2, max: 7) |
| Q5: [SC] What is the maximum DISPLAY mileage (km) of your BEVs when fully charged? | *Display Mileage* | - [250, 350) (n= 45, 20.7%)<br>- [350, 450) (n=119, 54.8%)<br>- [350, 450) (n=37, 17.1%)<br>- Over 550 (n=16, 7.4%) |
| Q6: [SC] What is the maximum REAL mileage (km) of your BEVs when fully charged? | *Real Mileage* | - [250, 350) (n= 48, 22.1%)<br>- [350, 450) (n=120, 55.3%)<br>- [350, 450) (n=39, 18%)<br>- Over 550 (n=10, 4.6%) |
| Q7: [SSC] For a highway trip that is beyond the real mileage of a BEV (i.e., you will need to recharge once in the middle of the trip). If the waiting time before charging is $t$ minutes, would you choose a BEV or a fuel car?<br>- $t$ = [0, 15, 30, 60]<br>- The *Comfort Time* is defined as the minimum $t$ when a participant chooses fuel car. If s/he never chooses a fuel car, *Comfort Time* is set as above 60 minutes. | *Comfort Time* | - 0 min (n= 30, 13.8%)<br>- 15 min (n=23, 10.6%)<br>- 30 min (n=41, 18.9%)<br>- 60 min (n=28, 12.9%)<br>- Above 60 min (n=95, 43.8%) |
| Q8: [SSC] If the trip is $m$ km and there are no charging stations along the way, what is your minimum comfortable percent of display mileage before the trip starts?<br>- $m$ = [25%, 50%, 75%] * *Display Mileage* | *Comfort Mileage 25%*<br>*Comfort Mileage 50%*<br>*Comfort Mileage 75%* | - mean: 39.7% (SD: 9.4, min: 27.5, max: 80)<br>- mean: 64.8% (SD: 9.5, min: 52.5, max: 100)<br>- mean: 85.7% (SD: 6.2, min: 77.5, max: 100) |
| Q9: [SSC] You are driving on highway. When approaching an upcoming rest area, the navigation informs you that the waiting time before charging at the area is $t$ minutes, the remaining battery range of the BEV is $r$ km and you are $d$ km away from destination (where you have plenty of time to recharge), would you choose to charge at this area or charge at the destination?<br>- $t$ = [0, 15, 30, 45, 60, Above 60]<br>- $r$ = [25%, 50%, 75%] * Display Mileage<br>- $d$ = [30%, 45%, 60%, 75%, 90%] * $r$ | *Waiting Time* ($t$)<br>*Rest Battery* ($r$)<br>*Rest Trip* ($d$)<br>*Charge Decision* | - [0, 15, 30, 60, Above 60] min<br>- [25%, 50%, 75%]<br>- [30%, 45%, 60%, 75%, 90%]<br>- Charge at the upcoming rest area (51.3%)<br>- Charge at the destination (48.7%) |

*Note: Abbreviations of question types are as follow: FI: Fill-in-text; SC: Single-choice; MC: Multiple-choice; LS: Likert scale, SSC: Scenario Single-choice; TF: True or false. SD standards for standard deviation.*

## METHOD

### Questionnaire Design

The designed questionnaire is illustrated in Table 1. Q1 to Q6 collect demographic and driving/vehicle-related information as they may affect drivers' decisions. In Q7, we defined *Comfort Time* to reveal the individual differences in tolerating waiting time before charging, as different BEV users may exhibit different levels of time-related anxiety. Q8 assessed distance-related anxiety. In Q9, we designed scenarios with three within-subjects factors (i.e., *Waiting Time, Rest Battery, and Rest Trip*) to reveal the tradeoff between distance-based and time-based anxiety. A Chinese version of the questionnaire was used, as this study was targeted toward BEV users in mainland China.

### Participants

All participants were recruited via social media on the Internet (e.g., Wechat and Weibo). A total of 344 participants completed the questionnaire. We then screened the answers based on two quality-checking questions (i.e., "please select the first/second option if you are reading the questionnaire") and removed answers from the drivers who drive BEV for commercial purposes (e.g., taxi drivers) as they may have developed different strategies when using BEVs. Finally, 217 samples were kept for analysis (Male: 159; Female: 58). These 217 drivers received a compensation of 5 RMB for their completion of the 15-minute-long questionnaire. This study was approved by the Human and Artefacts Research Ethics Committee at the Hong Kong University of Science and Technology (protocol number: HREP-2022-0051).

## Statistics Models

Two statistical models were built, in order to investigate: 1) influential factors of drivers' *Comfort Time* (i.e., Time Model); 2) how driver makes *Charge Decision* in the scenarios (i.e., Decision Model) when both distance-based and time-based factors matter. Both models were built in "SAS OnDemand for Academics". All other variables other than the dependent variables, as well as their two-way interactions were used as independent variables.

The Time Model was an ordered logistic regression model. We modeled the odds that a participant tolerates a longer waiting time versus a shorter waiting time before charging. For Decision Model, significant factors in Time Model were abandoned, and the rest variables were used as predictors in a binary logistic regression model. We modeled the odds of participants choosing to "*charge at the destination*" as compared to "*charging at the upcoming rest area*". Both Time Model and Decision Model were built with GENMOD procedure, and repeated measures were accounted for through a generalized estimating equation, which can be used to model multiple responses from a single subject.

In the models, the *Comfort Mileage*, *Waiting Time*, *Rest Battery*, *Rest Trip* were treated as continuous variables in the model as we care about the trends of their influence instead of the influence of specific factor levels. Before fitting the models, the correlations between all independent variables were assessed. The correlated independent variables were aggregated or abandoned, if possible, based on Quasi-likelihood under the Independence model Criterion (QIC) (W. Pan, 2001) of the models fitted with the variables. Thus, the variables in the full Time Model included *Age, Driving Frequency, Infrastructure, BEV Trust, Display Mileage, Real Mileage, and Comfort Mileage 50%*; and the variables in the full Decision Model included *Comfort Time*, the scenario variables in Q9 (i.e., *Waiting Time, Rest Battery, Rest Trip*), as well as all variables that were excluded in the Time Model. A backward model selection method was adopted based on QIC to select variables in both models. It should be noted that in the Decision Model, the scenario variables in Q9 were always kept in the model as we aim to model their influence on drivers' charging decisions. Post-hoc comparisons were made if the main effects or the two-way interactions of the independent variables were significant ($p < .05$).

## RESULTS

We report all significant effects ($p < .05$) in the fitted models after model selection, as well as the significant post-hoc contrasts. It should be noted that in the model selection process, all dropped variables were not significant ($p > .05$).

## Time Model

Table 2 summarizes the results of Time Model, in which, the users' trust in RES of BEVs (BEV Trust) and users' Comfort Mileage were found to be significant predictors of users' Comfort Time.

Specifically, it was found that with every 1-unit increase in *BEV Trust*, the odds of drivers tolerating longer waiting time increases by 40.6% (95%CI: 10.8%, 78.4%). At the same time, with every 10% increase in the *Comfort Mileage 50%* (i.e., the variable extracted from Q8), drivers are more likely to tolerate shorter waiting time, with an OR of 0.73, 95%CI: [0.57, 0.93]. It should be noted that *Comfort Mileage 50%* is correlated with *Comfort Mileage 25%* ($r = 0.72$, $p = <.0001$) and *Comfort Mileage 75%* ($r = 0.77$, $p = <.0001$); thus, only *Comfort Mileage 50%* was kept in the model.

Table 2. Summary of Time Model Results

| Independent Variable | $\chi^2$-value | *p*-value |
|---|---|---|
| BEV Trust | $\chi^2(1) = 7.86$ | .005** |
| Driving Frequency | $\chi^2(1) = 3.61$ | .06* |
| Comfort Mileage 50% | $\chi^2(1) = 6.74$ | .01** |

*Notes: In this table and the following tables, * marks marginal significant predictors (.05<p<.1), ** marks significant predictors (p<.05); IV stands for Independent Variable.*

## Decision Model

As shown in Table 3, five significant two-way interaction effects for *Charge Decision* have been observed.

*Interaction between Comfort Time and Waiting Time*. As shown in Figure 1a, in general, with the increase of the *Waiting Time* at the upcoming rest area, drivers become more likely to charge at the destination, but to different extents among drivers who reported different lengths of *Comfort Time*. Specifically, for every 10-minute increase in waiting time at the upcoming rest area, the ORs of drivers who reported 30-minute and 60-minute *Comfort Time* to charge at the destinations are 1.15 (95%CI: [1.05, 1.26], $\chi^2(1) = 8.84$, $p=.003$) and 1.11 (95%CI: [1.02, 1.21], $\chi^2(1) = 5.23$, $p=.02$).

Table 3. Summary of Decision Model Results

| Independent Variable | $\chi^2$-value | *p*-value |
|---|---|---|
| Real Mileage | $\chi^2(3) = 4.97$ | .2 |
| Infrastructure | $\chi^2(2) = 1.15$ | .6 |
| Comfort Time | $\chi^2(4) = 7.27$ | .12 |
| Rest Battery | $\chi^2(1) = 56.38$ | <.0001** |
| Rest Trip | $\chi^2(1) = 71.20$ | <.0001** |
| Waiting Time | $\chi^2(1) = 2.83$ | .09* |
| Comfort Time * Waiting Time | $\chi^2(4) = 18.54$ | .001** |
| Infrastructure * Rest Trip | $\chi^2(2) = 6.67$ | .04** |
| Infrastructure * Rest Battery | $\chi^2(2) = 8.60$ | .014** |
| Rest Battery * Waiting Time | $\chi^2(1) = 5.21$ | .02** |
| Rest Trip * Waiting Time | $\chi^2(1) = 20.04$ | <.0001** |

*Interaction between Infrastructure and Rest Trip*. As shown in Figure 1b, in general, with the increase of the *Rest Trip*, drivers become less likely to charge at the destination, but to different extents in regions with different levels of EV infrastructure. Specifically, with every 10% increase in *Rest Trip*, for drivers from the regions with less-developed EV infrastructure, average EV infrastructure, and well-developed EV infrastructure, their odds to charge at the destinations are estimated to decrease by 9.5% (95%CI: [4.1%, 14.6%], $\chi^2(1) = 11.54$, $p = .0007$), 17.1% (95%CI: [12.3%, 21.5%], $\chi^2(1) = 43.85$, $p < .0001$), and 16.4% (95%CI: [12.2%, 20.4%], $\chi^2(1) = 51.24$, $p < .0001$).

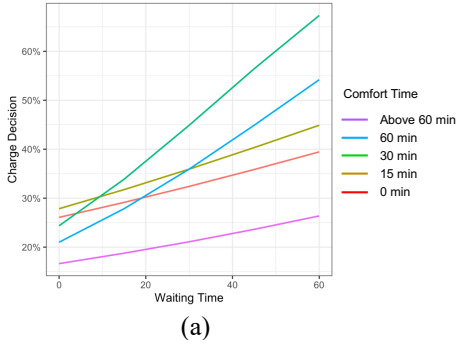

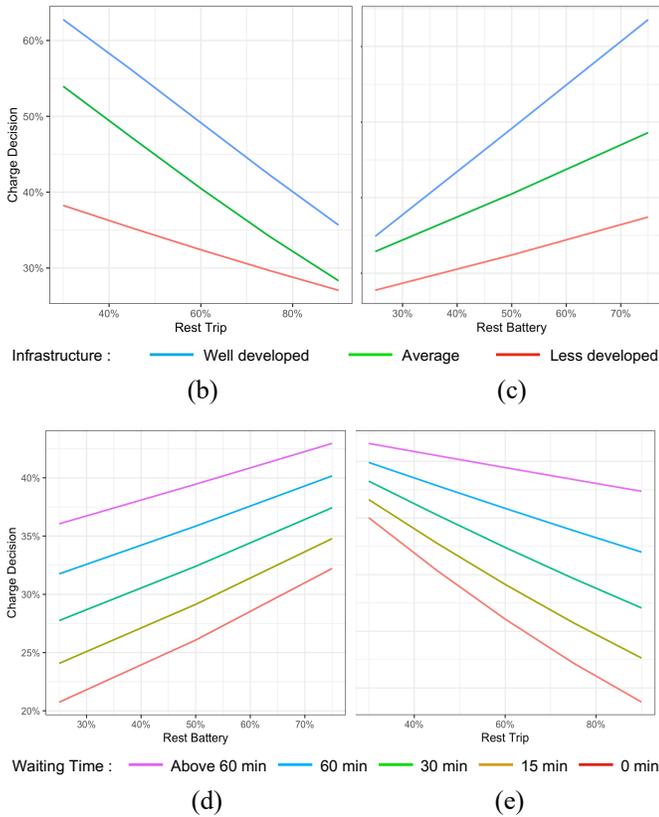

Figure 1. Visualization of significant interaction effects in Decision Model. The vertical axis indicates the predicted probability (from 0% to 100%) of choosing to charge at the destination.

*Interaction between Infrastructure and Rest Battery*. As shown in Figure 1c, EV infrastructure development also interacts with *Rest Battery* of the BEVs, i.e., with the increase of *Rest Battery*, drivers are more likely to charge at the destination, but again, to different extents in regions with different levels of the EV infrastructure (less-developed: OR=1.09, 95%CI: [1.03, 1.14], $\chi^2(1) = 10.54$, $p = .001$; average: OR=1.14, 95%CI: [1.08, 1.20], $\chi^2(1) = 23.20$, $p < .0001$; developed: OR=1.22, 95%CI: [1.14, 1.30], $\chi^2(1) = 37.94$, $p < .0001$).

*Interaction between Rest Battery and Waiting Time*. As shown in Figure 1d. It seems that with the increase in the *Waiting Time*, drivers become less sensitive to the increase of the *Rest Battery*. In general, with the increase of the *Rest Battery*, drivers become more likely to charge at the destination, but to different extents with different lengths of waiting time. For example, when the *Waiting Time* is 0, 15, 30, and 60 minutes, for every 10% increase of the *Rest Battery*, the ORs of charging at the destination is 1.15 (95%CI: [1.11, 1.19], $\chi^2(1) = 56.38$, $p < .0001$), 1.13 (95%CI: [1.10, 1.17], $\chi^2(1) = 60.47$, $p < .0001$), 1.12 (95%CI: [1.09, 1.15], $\chi^2(1) = 55.03$, $p < .0001$) and 1.10 (95%CI: [1.06, 1.14], $\chi^2(1) = 25.23$, $p < .0001$).

*Interaction between Rest Trip and Waiting Time*. As shown in Figure 1e, with the increase in the *Waiting Time*, drivers also become less sensitive to the increase in the *Rest Trip*. When the *Waiting Time* is 0, 15, 30, and 60 minutes, for every 10% increase in the *Rest Trip*, the ORs of charging at the destination is 0.86 (95%CI: [0.83, 0.89], $\chi^2(1) = 71.20$, $p < .0001$), 0.87 (95%CI: [0.85, 0.90], $\chi^2(1) = 70.95$, $p < .0001$), 0.89 (95%CI: [0.87, 0.92], $\chi^2(1) = 60.31$, $p < .0001$) and 0.93 (95%CI: [0.90, 0.96], $\chi^2(1) = 19.82$, $p < .0001$).

## DISCUSSION

In this study, through an online questionnaire, we conducted a preliminary investigation on BEV users' acceptance of waiting time at the charging stations (i.e., *Comfort Time*) and how waiting time at the charging station can impact BEV users' charging decisions when both the time cost and distance-related range anxiety are of concern.

Through Time Model, we modeled individual differences in tolerating waiting time when charging (i.e., *Comfort Time*). It was found that those who trust more in the range estimation system of BEVs (*BEV Trust*) and those who can accept a lower battery level before a trip (*Comfort Mileage 50%*) can tolerate longer waiting time. It is not difficult to understand the association between *Comfort Mileage* and *Comfort Time*: those who prefer a higher battery level at the beginning of the trip might be those who wish to avoid charging in the middle of a trip to save time. As for the relationship between the *BEV Trust* and *Comfort Time*, we assume there might be covariates influencing both of them (e.g., personalities, Yuan et al., 2018), and it deserves further investigation, but it is beyond the scope of this survey.

At the same time, as a major contribution of this study, we observed that BEV users' charging decisions could be affected by time-related factors, and this influence can be moderated by a number of other factors. In general, as expected, we have observed that with the increase in the *Waiting Time* at the upcoming rest area, drivers tended to skip this rest area and charge at the destination. This trend indicates that the "range anxiety" as well as drivers' charging decisions are not solely affected by "whether I can reach the destination" but also "how much time it will take for me to arrive." Future research should take "time" into consideration when optimizing the distributions of charging infrastructures in addition to "distance" (e.g., Hafez & Bhattacharya, 2017; L. Pan et al., 2020) in order to optimize BEV users' experience. However, it should be noted that our study only collected users' responses in imagined scenarios. Future research may need to further validate the relationship we have observed in more realistic experiment setups (e.g., naturalistic driving studies or observational studies).

Further, the influence of *Waiting Time* on users' charging decisions can also be moderated by *Rest Battery* and *Rest Trip*. Especially, these interaction effects revealed how drivers traded off between distance-related anxiety (as materialized by *Rest Battery* and *Rest Trip*) and time-related anxiety (i.e., *Waiting Time*). For example, as expected, with the decrease of the Rest Battery, drivers are more inclined to charge at the upcoming rest area to reduce the risk of running out of battery before arrival. However, when the waiting time at the upcoming rest area becomes longer, drivers become less sensitive to the battery level. Similar trends have also been observed for *Rest Trip* – without considering the waiting time, with the increase of the *Rest Trip*, drivers prefer to charge at the upcoming rest area to reduce the uncertainty in the rest of the trip. However, a longer waiting time leads to less sensitivity in the *Rest Trip* (i.e., drivers would still decide to "take risk" and finish the trip if the waiting time is long). For future work, we should be able to identify decision margins considering distance-related and time-related range anxiety, which can support the charging network optimization and charging recommendation systems.

Lastly, we also observed regional differences when BEV drivers made decisions. The influence of both *Rest Trip* and *Rest Battery* becomes less obvious in regions with less developed EV infrastructure compared to those regions with more developed EV infrastructure. Drivers also tend to be more inclined to charge at the upcoming rest areas in less developed regions. This is a novel finding, but it is not surprising – drivers from regions with less developed EV infrastructure may have higher anxiety about running out of battery, and charging their car whenever possible can be a less risky decision. Further, it should be noted that the natural environment (e.g., temperature, Wang et al., 2023) may also have affected BEV users' charging habits and should be considered in future studies.

## CONCLUSIONS

In this work, through a scenario-based survey study, we disentangled range anxiety into the distance- and time-related part. Based on the results, we found that users' trust in RES of BEVs can influence users' acceptable waiting time. Further, our results showed that users' charging decisions can be affected by both time-related and distance-related range anxiety, which can be further moderated by several scenario-related factors. These findings suggest that to improve user experience, the charging recommend systems and the optimization of the charging stations should take charging time and users' characristics into consideration. Future studies should explore the influence of more social-psychological factors on charging decisions and extract computable models based on a larger dataset.

## Acknowledgement

This work was supported by the Guangzhou Municipal Science and Technology Project (No. 2023A03J0011) and the Guangzhou Science and Technology Program City-University Joint Funding Project (No. 2023A03J0001).